\documentclass[12pt]{article}
\usepackage{epsfig}
\usepackage{amsmath}
\usepackage{amsfonts}

\begin{document}
\pagestyle{empty}
\begin{flushright}
UMN-TH-2527/06\\
November 2006
\end{flushright}
\vspace*{5mm}

\begin{center}

{\Large\bf Yang-Mills Localization in Warped Space}
\vspace{1.0cm}

{\sc Brian Batell}\footnote{E-mail:  batell@physics.umn.edu}
{\small and} 
{\sc Tony Gherghetta}\footnote{E-mail:  tgher@physics.umn.edu}
\\
\vspace{.5cm}
{\it\small {School of Physics and Astronomy\\
University of Minnesota\\
Minneapolis, MN 55455, USA}}\\
\end{center}

\vspace{1cm}
\begin{abstract}
We present a mechanism to localize zero mode non-Abelian gauge fields in a slice of AdS$_5$. As in the U(1) case, bulk and boundary mass terms allow for a massless mode with an exponential profile that can be localized anywhere in the bulk. However in the non-Abelian extension, the cubic and quartic zero mode gauge couplings do not match, implying a loss of 4D gauge invariance. We show that the symmetry can be restored at the nonlinear level by considering brane-localized interactions, which are added in a gauge invariant way using boundary kinetic terms. Possible issues related to the scalar sector of the theory, such as strong coupling and ghosts, are also discussed. Our approach is then compared with other localization mechanisms motivated by dilaton gravity and deconstruction. Finally, we show how to localize the scalar component $A_5$ zero mode anywhere in the bulk which could be relevant in gauge-Higgs unification models.
\end{abstract}

\vfill
\begin{flushleft}
\end{flushleft}
\eject
\pagestyle{empty}
\setcounter{page}{1}
\setcounter{footnote}{0}
\pagestyle{plain}

\section{Introduction}
Gauge theories in a slice of anti-de Sitter (AdS) space have attracted a great deal of attention in recent years primarily because they provide geometrical explanations of particle physics mysteries, including the gauge hierarchy problem \cite{rs}, and a window into strongly coupled gauge theories via the AdS/CFT correspondence \cite{ads1,ads2,ads3,pheno1,pheno2,pheno3,pheno4}. Massless gauge fields propagating in the bulk of AdS$_5$ have been thoroughly examined in the literature \cite{gaugeb1, gaugeb2} and have many applications (see for example~\cite{review}). It is well-known that the zero mode of this field has a flat profile (constant wavefunction) in the extra-dimension, and that the dual theory is described by a (mostly) elementary source field interacting with a strongly coupled conformal field theory (CFT) via a conserved CFT current \cite{pheno1, pheno4}.   

Not so familiar is the fact that gauge fields can be localized in the extra dimension, at least for the Abelian case \cite{gbl1,gbl2}. In order to localize the zero mode, the bulk equation of motion and boundary conditions must be modified, which is accomplished through the introduction of mass terms in the bulk and on the boundaries. By tuning these masses, a zero mode solution with an exponential profile is allowed. The phenomenology and holographic description of localized U(1) gauge fields is described in \cite{u1}. 

Extending this mechanism to the non-Abelian case is not so straightforward. In order to preserve four-dimensional (4D) gauge invariance, the zero mode cubic and quartic gauge interactions must have the same coupling. These couplings are proportional to wavefunction overlap integrals, and for exponential zero mode wavefunctions they do not necessarily match. Thus it seems the localization mechanism violates 4D gauge invariance. 

To have a viable low energy gauge theory, the equality of these couplings must be restored. In this paper we show that this can be achieved by considering interactions on the boundary, which provide the correct contributions to zero mode cubic and quartic terms to force the gauge couplings to be equal. These interactions can be added to the theory in a gauge invariant way via boundary kinetic terms.    

Localizing both Abelian and non-Abelian gauge fields is important for both 5D model building and phenomenology as well as holographic physics. Because the gauge field can be localized anywhere in the bulk, an entire new range of parameter space can now be explored. For example, if the Standard Model gauge bosons all propagate in the bulk, each field could in principle be localized at different places in the extra dimension. This may help to alleviate experimental bounds, as in the U(1) case, and could perhaps lead to novel model building scenarios. Perhaps more interesting is the notion that if zero modes are localized  near the infrared boundary, there may exist a dual theory in which the gauge fields are composite CFT states. In this case, the dual gauge theory would be a valid effective field theory below the cutoff corresponding to the scale of conformal symmetry breaking. 

Our approach is not without a few caveats that need to be addressed. First, a particular boundary kinetic term must be added to restore the gauge invariance at the nonlinear level. This represents an additional tuning beyond the tuning already required between bulk and boundary masses (as well as the tuning between bulk and boundary cosmological constants in the original Randall-Sundrum model!). More importantly, mass terms in the 5D theory must be generated through spontaneous symmetry breaking if we hope to preserve a 4D gauge symmetry. Therefore, we must also account for the additional scalar fields present in the theory. These scalar fields complicate the analysis and may also introduce more severe problems, such as ghosts or strong coupling. Although we are not able to offer a general analysis of these scalar issues, we present some simple examples illustrating the problems and speculate on possible resolutions.  Even if these issues can only be resolved in the UV completion of the model the nice features of the Yang-Mills localization will still be preserved at low energies and therefore makes the present study relevant.
 
We also compare our approach to other methods of localizing gauge fields on the brane. As first shown in \cite{dilaton1}, gauge fields can be given a localized wavefunction through a dilatonic coupling. A similar mechanism is also motivated by deconstruction, in which the 4D gauge theories have nonuniversal gauge couplings at each site \cite{decon}. These approaches share many of the features of our method, but there are also differences, which we point out. 

The scalar component $A_5$ is also utilized in models of electroweak symmetry breaking as an alternative to the Higgs boson \cite{hosotani}. The zero mode profile of this field is restricted by gauge invariance. However, we show that like the vector $A_\mu$, it is possible to change the profile, and thus localize the $A_5$ zero mode anywhere in the bulk. This could affect the phenomenology of
gauge-Higgs unification models in novel ways.

The organization of this paper is as follows. In Section 2 we begin by reviewing how the addition of bulk and boundary mass terms allow for a localized zero mode gauge field. We then move on to the full non-Abelian analysis, showing how boundary kinetic terms, if appropriately tuned, can restore 4D gauge invariance. In Section 3 we discuss the additional scalar fields in the theory. We present a flat space analysis of the Nambu-Goldstone bosons associated with the massive vectors in the 4D theory. We also discuss the possibility of strong coupling and ghosts by considering a simple example containing only boundary masses (no bulk masses). In Section 4 we review other methods of gauge field localization, including localization with a dilaton and via deconstruction, and compare these methods with our approach. We discuss the localization of the scalar $A_5$ in Section 5. Finally we conclude in Section 6, and suggest possible directions for future investigation.

\section{Non-Abelian gauge fields in warped space}
We will work with the following metric describing 5D anti-de Sitter space:
\begin{equation}
ds^2=e^{-2 ky}\eta_{\mu\nu}dx^{\mu} dx^{\nu}+dy^2~,
\label{adsmetric}
\end{equation}
where $y$ is the extra coordinate and $k$ is the curvature scale related to the bulk cosmological constant. The fifth dimension $y$ is compactified on a $Z_2$ orbifold and two three-branes are located at the orbifold fixed points $y=0$ and $y=\pi R$, referred to as the ultraviolet (UV) and infrared (IR) brane, respectively. The indices of 5D coordinates are labeled with Latin letters ($A$,$B$, $\dots$) while 4D indices are labeled by Greek letters ($\mu$, $\nu$, $\dots$). All 4D indices are raised and lowered with the Minkowski metric $\eta={\rm diag}(-~+~+~+)$.

It is well-known that massless gauge fields propagating on the background (\ref{adsmetric}) lead to a 
massless zero mode that is not localized in the extra dimension, but rather has a flat profile \cite{gaugeb1, gaugeb2}.  Instead to localize a zero mode, bulk and boundary mass terms must be added which modify both the bulk equation of motion and the boundary conditions for the gauge field \cite{gbl1, gbl2}. The massless mode then has an exponential profile and can be localized anywhere in the bulk. We will show how this is accomplished in Section 2.1. Of course such mass terms must be generated in a gauge invariant way, so that a 4D gauge symmetry is preserved for the zero mode. The most obvious way to generate such terms is by the spontaneous breaking of gauge symmetry. This introduces extra scalar fields into the theory which may cause problems. For instance, if massless scalar modes persist in the 4D theory they may be undesirable for model building. Other, more severe problems, such as the existence of ghosts or strongly coupled light scalar fields, may also occur. We will discuss these issues in Section 3. However, the gauge sector can be isolated from the scalar sector, and none of these more severe problems exist for the gauge field.

A separate issue arises when we consider the localization of non-Abelian gauge fields. As we will see in this section, the zero mode cubic and quartic couplings do not match, and thus the conventional 4D gauge symmetry does not exist in the simplest formulation. The effective couplings are proportional to the overlap integral of the wavefunctions, and are only equal when the zero mode is flat. To remedy this, we will add cubic and quartic interactions on the branes which will restore the equality of the zero mode gauge couplings. The only way this can be done while preserving gauge invariance is by adding a kinetic term to the brane, which we show in Section 2.2. We study only pure Yang-Mills theory in this paper. However, if fermions propagate in the bulk, we can generalize the mechanism and restore 4D gauge invariance by adding gauge covariant boundary kinetic terms for fermions.

\subsection{Review of the localization mechanism}

We begin by briefly reviewing how to localize the zero mode gauge field in the extra dimension. For simplicity we will consider a U(1) gauge field, since the generalization to the non-Abelian case is straightforward at the quadratic level in the action. Consider the following 5D equation of motion for a spin-1 field $A_M(x,y)$:
\begin{equation}
\frac{1}{\sqrt{-g}}\partial_M\left(\sqrt{-g}F^{MN}\right)-M^2(y)A^N=0~.
\label{eom1}
\end{equation}
The mass term arises from spontaneous symmetry breaking \cite{gbl2, gbmass} and contains both bulk and boundary terms: 
\begin{equation}
M^2(y)=ak^2+ \alpha \sigma''(y)~,
\label{mass}
\end{equation}
where we have defined $\sigma''(y)=2k(\delta(y)-\delta(y-\pi R))$. The bulk and boundary masses are written in terms of the AdS curvature $k$ with dimensionless parameters $a$ and $\alpha$. As shown in \cite{gbl1,gbl2, u1}, we must tune the parameters,
\begin{equation}
\alpha=1\pm\sqrt{1+a}~,
\label{tune}
\end{equation}
to allow a zero mode solution. Only real values for the mass parameters, $-1<a<\infty$, are considered and therefore $\alpha$ can take any real value. 

Let us choose the gauge $A_5=0$ for simplicity. The equation of motion (\ref{eom1}) becomes
\begin{equation}
\left[e^{2ky}\Box + \partial^2_5 -2k\partial_5 -a k^2
-  \alpha\sigma''(y)\right]A_\mu(x,y)=0~.
\label{eom2}
\end{equation}
Decomposing the vector $A_\mu$ as
\begin{equation}  
A_\mu(x,y)=\sum_{n=0}^\infty A_\mu^n(x) f^n(y)~,
\label{kk}
\end{equation}
the solution of (\ref{eom2}) for the massless mode is
\begin{equation}
f^0(y)\sim e^{\alpha k y}~.
\end{equation}
Since $\alpha$ can be any real number (\ref{tune}), the zero mode can be localized anywhere in the fifth dimension.

To see the necessity of tuning the parameters $a$ and $\alpha$, note the boundary condition derived by integrating the equation of motion (\ref{eom2}) across the boundary is 
\begin{equation}
(\partial_5 -\alpha k)f^n(y)\bigg\vert_{0,\pi R}=0~.
\label{bc1} 
\end{equation}
The zero mode satisfies this equation only if the relation (\ref{tune}) holds. Otherwise, the boundary conditions project out the massless mode. It is possible that this tuning is enforced in a supersymmetric extension, as happens for bulk scalar fields \cite{gp1}. 

To summarize, the bulk mass term changes the profile of the zero mode from flat to exponential. Clearly, an exponential wavefunction is incompatible with ordinary Dirichlet or Neumann conditions. The brane-localized mass modifies the boundary condition to be compatible with this exponential wavefunction. 
Furthermore, this leads to interesting phenomenological possibilities and remarkably the localized U(1) gauge bosons can be given a holographic interpretation~\cite{u1}. 

\subsection{Non-Abelian case}
Let us now consider the following action for a non-Abelian gauge field $A_M^a$ in a slice of AdS$_5$:
\begin{eqnarray}
S&=&\int d^5 x \sqrt{-g}  \left[-\frac{1}{4}F^a_{MN}F^{MNa}-\frac{1}{2}M^2(y)A^a_M A^{Ma}\right]~,
\label{a1}
\end{eqnarray}
where the field strength tensor is defined as 
\begin{equation}
F^a_{MN}=\partial_M A^a_N-\partial_N A^a_M + g_5 f^{abc} A^b_M A^c_N~,
\end{equation}
with group structure constants $f^{abc}$ and 5D coupling $g_5$. Again, the mass term is assumed to be generated by spontaneous symmetry breaking. 

The technique described in the previous section can also be applied to localizing non-Abelian gauge fields. At the quadratic level, the equation of motion is the same as in the Abelian case, and therefore the procedure is identical. However, when we examine the nonlinear terms of the Yang-Mills theory to compute the zero mode cubic and quartic interactions, we find that the couplings do not match. Thus it seems that there is no longer any 4D gauge invariance.

It is possible to restore the zero mode cubic and quartic couplings consistent with 4D gauge invariance if we consider brane-localized interactions. These terms play a similar role to the boundary mass terms, which cancels the bulk mass contribution to the zero mode action. We cannot add these terms by hand to the boundary action without violating gauge invariance. To preserve the symmetry, we must add a boundary kinetic term, so that the action becomes
\begin{equation}
\int d^5x \left[-\frac{1}{4}\left(1+\frac{\beta}{k^2}\sigma''(y)\right)F^a_{\mu\nu}F^{\mu\nu a}+\dots\right].
\label{a2}
\end{equation}
Note the boundary term is suppressed by $k^{-1}$ relative to the bulk term. It will be necessary to tune the dimensionless coefficient $\beta$ to match the zero mode couplings. For simplicity, we have chosen the same coupling on each brane, but it is possible to have them different, which we will discuss at the end of this section.

Also notice that one of the kinetic terms has a negative coefficient and therefore could introduce a ghost into the theory. However, it is the sum of the bulk and boundary kinetic energies which is important. When we perform the Kaluza-Klein reduction, we will require the wavefunction overlap integral, including the bulk and boundary terms, to be positive, which then implies the particle has positive energy. We will see that this requirement is satisfied for all values of $\beta$ that we will consider.  In fact negative boundary kinetic terms have been previously considered in the literature. In Ref. \cite{blkt2} the effect of brane kinetic terms in the Randall-Sundrum model was studied and it was found that a ghost does indeed appear in the spectrum, but only if one of the boundary terms is made so strong and negative that it overwhelms the positive energy contributions from the bulk and other boundary. Similar conclusions were drawn in Ref. \cite{blkt3} where negative kinetic terms in flat space were examined. Thus, it seems from the 4D effective field theory vantage point, negative boundary kinetic terms in principle pose no problem to model building. However, note that this cancellation mechanism is nonlocal since locally there is always a negative kinetic term.

With this modification to the action, the equation of motion at quadratic level is
\begin{equation}
\left[e^{2ky}\left(1+\frac{\beta}{k^2}\sigma''(y)\right)\Box + \partial^2_5 -2 k \partial_5 - a k^2
-  \alpha  \sigma''(y) \right]A_\mu(x,y)=0~.
\label{eom4}
\end{equation}
Decomposing the field (\ref{kk}) and demanding 
\begin{equation}
\int_0^{\pi R} dy \left(1+\frac{\beta}{k^2}\sigma''(y)\right) f^n f^m=\delta^{nm}~,
\label{normal}
\end{equation}
the equation of motion for the eigenfunction $f^n(y)$ becomes
\begin{equation}
\left[\partial^2_5 -2 k \partial_5 - a k^2 
-  \alpha  \sigma''(y) \right]f^n(y)=-e^{2ky}m_n^2\left(1+\frac{\beta}{k^2}\sigma''(y)\right)~.
\label{eom5}
\end{equation}
The bulk equation of motion is unaffected by the presence of the boundary term, but the boundary condition is now altered:
\begin{equation}
\left(\partial_5 -\alpha k+\frac{\beta}{k} m_n^2e^{2ky}\right)f^n(y)\bigg\vert_{0,\pi R}=0~.
\label{bc2}
\end{equation}
The massive modes are influenced by the brane kinetic term and thus the spectrum will be modified. However, the boundary condition for the massless mode $f^0(y)$ is the same as in the Abelian case 
(\ref{bc1}), and therefore the zero mode can be localized provided we demand the tuning condition (\ref{tune}). The normalized zero mode solution is given by
\begin{equation}
f^0(y)=\sqrt{\left(\frac{1}{1-2\alpha \beta}\right)\left(\frac{2\alpha k}{e^{2\alpha \pi k R}-1}\right)}~e^{\alpha k y}\equiv N e^{\alpha k y}~.
\label{wf}
\end{equation}
Note that for a flat zero mode ($\alpha=0$), the normalization is unaffected by the presence of the brane-localized kinetic term since the wavefunction has the same value at each boundary. The normalization $N$ must be real, or equivalently the overlap integral (\ref{normal}) must be positive so that the field is not ghostlike. Thus, we must require that $\beta \alpha<1/2$. We will see shortly that this limit is always satisfied if we require that a 4D gauge invariance be preserved for the zero mode.

\subsubsection{Cubic and quartic couplings}

The interaction terms following from the action (\ref{a2}) are
\begin{equation}
S_{int}=- \int d^5x~\left(1+\frac{\beta}{k^2}\sigma''(y)\right)\left[g_5 f^{abc}\partial_\mu A^a_\nu A^{\mu b}A^{\nu c}+\frac{g_5^2}{4}f^{abc}f^{ade}A^b_\mu A^c_\nu A^{\mu d}A^{\nu e}\right]~.
\label{int1}
\end{equation}
Localizing a massless mode comes at a price: the loss of {\it 4D gauge invariance} of the zero mode. Without boundary interaction terms, the cubic and quartic vertices will have different strengths. The 5D symmetry may still be realized in the effective theory, but it will not take the standard form familiar in 4D theories. The gauge invariance can be restored if we add a boundary kinetic term with a precise coefficient (\ref{a2}). 

Let us investigate the effective 4D interactions and see the problem and its resolution explicitly. Inserting the zero mode wavefunction (\ref{wf}) into (\ref{int1}), the couplings can be calculated by performing the overlap integrals. The cubic coupling is given by:
\begin{eqnarray}
g_3& =&  g_5 N^3\int_0^{\pi R}dy~ \left(1+\frac{\beta}{k^2}\sigma''(y)\right) ~ e^{3 \alpha k y}~,
 \nonumber \\
& &= \frac{g_5 N^3}{k}\left(\frac{1}{3 \alpha}-\beta\right)\left( e^{3\alpha \pi k R}-1 \right)~,
\label{cubic}
\end{eqnarray}
while the quartic coupling is 
\begin{eqnarray}
g_4^2&=&g_5^2N^4\int_0^{\pi R} dy ~ \left(1+\frac{\beta}{k^2}\sigma''(y)\right)~ e^{4\alpha k y}~,\nonumber\\
& =&\frac{g_5^2N^4}{k}\left(\frac{1}{4 \alpha}-\beta\right)\left( e^{4\alpha \pi k R}-1 \right)~.
\label{quartic}
\end{eqnarray} 
Clearly for $\beta=0$, $g_3\neq g_4$ and the 4D gauge symmetry is lost 
for general values of $\alpha$. Thus we see the necessity of including a boundary kinetic term for nonzero $\alpha$. Only when the mode is flat ($\alpha=0$) are the couplings identical. Moreover, when the mode is flat arbitrary boundary kinetic terms (with different coefficients) may be added to the action without affecting the zero mode gauge invariance.
 
We will demand that a 4D gauge symmetry be present in the theory for general values of $\alpha$, or in other words, we will require that $g_3=g_4$. To find the value of $\beta$ that satisfies this requirement, we square (\ref{cubic}) and equate it to (\ref{quartic}). Solving for $\beta$, we find
\begin{equation}
\beta_\pm=\frac{1}{3\alpha}\frac{1}{1-\omega}\left[1-\frac{9}{8}\omega\pm \frac{3}{8}\sqrt{\omega^2-\frac{8}{9}\omega}\right],
\label{beta}
\end{equation}
where we have defined
\begin{equation}
\omega=\frac{(e^{4\alpha \pi k R}-1)(e^{2\alpha \pi k R}-1)}{(e^{3\alpha \pi k R}-1)^2}
=1-\frac{1}{(1+2 \cosh[\alpha\pi k R])^2}~.
\end{equation}
The exponential modulating function $\omega$ ranges from as low as  $8/9$ when $\alpha=0$ and as high as $1$, when $|\alpha|$ is very large. 

\begin{figure}
\centerline{\includegraphics[width=.85\textwidth]{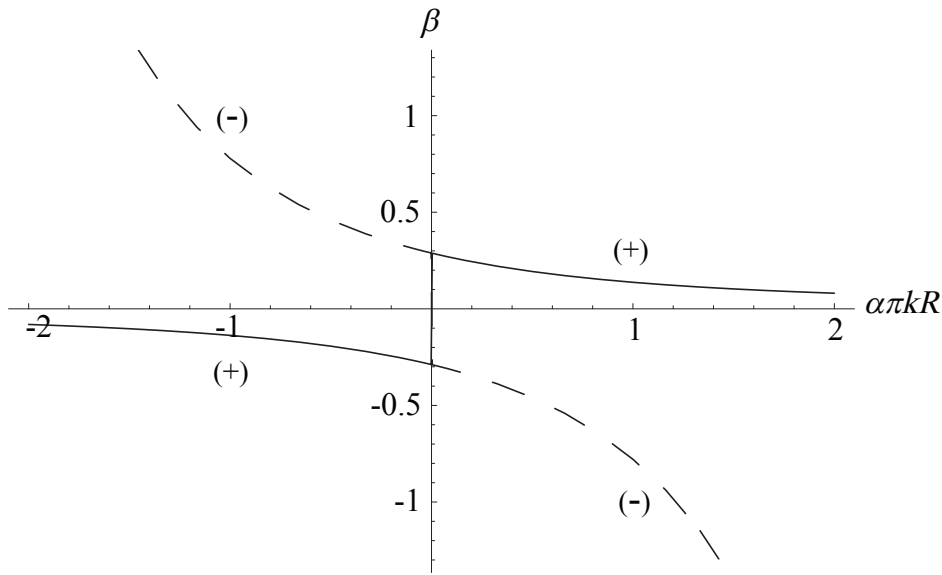}}
\caption{The boundary kinetic parameter $\beta_\pm$ (in units of $\pi k R$) as a function of 
$\alpha \pi k R$. The $\beta_+$ branch is indicated by the solid line while the $\beta_-$ branch is indicated by the dashed line. The $+$ and $-$ branches are continuously connected at $\alpha=0$ corresponding to a constant zero mode wavefunction. At $\alpha=0$, $\beta$ can take any value.}
\label{fig1}
\end{figure}

For any particular value of the localization parameter $\alpha$, there are two possible boundary kinetic terms that will preserve 4D gauge invariance, with strength $\beta_+$ or $\beta_-$. The $\beta_+$ and $\beta_-$ branches are continuously connected at $\alpha=0$ and this behavior is plotted in Fig.~\ref{fig1}.
It is instructive to study two limiting cases. For small values of $\alpha$, we find that the coefficient 
$\beta$ becomes
\begin{equation}
\beta_{\pm}=\pm \frac{1}{2\sqrt{3}}\pi k R +{\cal O}(\alpha)~.
\end{equation}
This equation seems to imply that for the flat mode ($\alpha=0$), we must also include boundary kinetic terms to maintain 4D gauge invariance. Of course this is not true. The presence of boundary kinetic terms has no effect on the 4D gauge invariance when the mode is flat, which can easily be seen, for example, from the wavefunction overlap integrals that determine the 4D couplings, Eqs. (\ref{cubic}) and (\ref{quartic}). The boundary terms simply cancel when the wavefunction is constant. In the more generic case of different coefficients for each brane-localized kinetic term, the zero mode couplings are still equal. 

We can also study the limit where $|\alpha|>0$, in which case the coefficient $\beta$ reduces to
\begin{eqnarray}
\beta_+&\simeq&\frac{1}{6\alpha}~,\label{beta-largealpha}\\
\beta_-&\simeq&-\frac{1}{12\alpha}e^{2|\alpha|\pi k R}~. \label{beta+largealpha}
\end{eqnarray}
We see that either a very small or very large boundary kinetic term must be added to restore gauge invariance in the low energy theory.  

\begin{figure}
\centerline{\includegraphics[width=.85\textwidth]{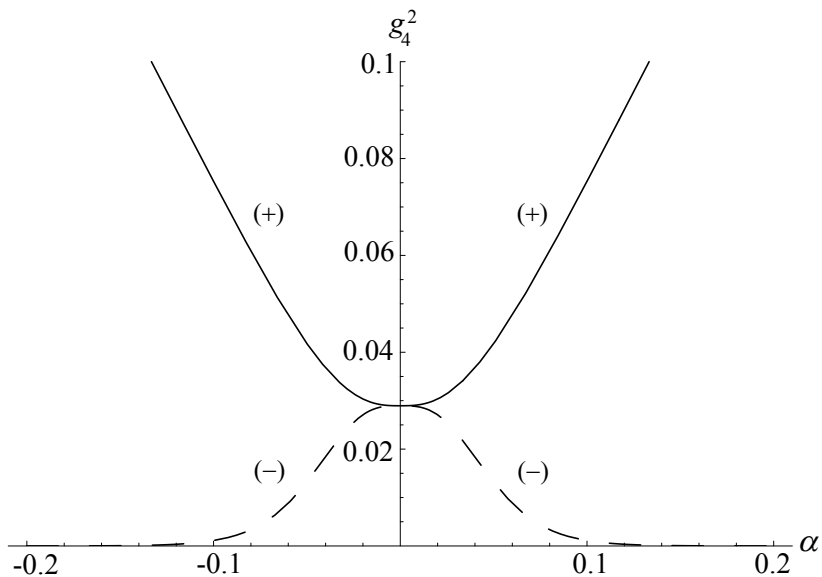}}
\caption{The 4D zero mode gauge coupling $g_4^2$ plotted (in units of $g_5^2 k$) as a function of $\alpha$ for $\pi k R =34.5$. For any given value of $\alpha$ there are two characteristic behaviors, the $\beta_-$ branch (dashed line), and the $\beta_+$ branch (solid line). The $+$ and $-$ branches are continuously connected at $\alpha=0$.}   
\label{fig2}
\end{figure}

Now we turn to the effective 4D gauge couplings, which are computed by inserting (\ref{beta}) back into (\ref{quartic}). The couplings can also take two possible values depending on which boundary kinetic term ($\beta_+$ or $\beta_-$) is chosen. This is shown clearly in Fig.~\ref{fig2} with $g_4^2$ plotted as a function of $\alpha$ in units of $g_5^2 k$.
The two branches are again connected at $\alpha=0$, in which case the 4D gauge coupling becomes the familiar $g_4^2=g_5^2/\pi R$. For large $|\alpha|$ the gauge coupling on the $\beta_-$ branch is approximately
\begin{equation}
g_{4-}^2\simeq 12|\alpha| (g_5^2 k)e^{-2|\alpha| \pi k R} ~.
\end{equation}
Notice that a large value of $g_5^2k$ is required to obtain a 4D gauge coupling of order one and therefore perturbativity of the 5D gauge theory does not allow for arbitrary localization on the 
$\beta_-$ branch. On the other hand for the $\beta_+$ branch (shown as the solid line in Fig.~\ref{fig2}), the gauge coupling for highly localized modes ($|\alpha|\gg0$) becomes:
\begin{equation}
g_{4+}^2\simeq \frac{3}{4} |\alpha| (g_5^2 k)~.
\end{equation}
In this case a 4D gauge coupling of order one for large $|\alpha|$ can always be obtained for
perturbative values of the 5D gauge coupling. Thus, for extremely localized zero modes, only the 
$\beta_+$ branch is consistent with perturbativity  for all large values of  $|\alpha|$.
For weak localization where $\alpha\sim0$, a 4D coupling of order one can be defined for both branches. As $|\alpha|$ becomes of order unity and the mode is localized, only the $\beta_+$ branch would be consistent with perturbative values of $g_5^2 k$.

Finally, let us verify that the zero mode is not a ghost for all values of $\beta$. At the end of Section 2.2 we found that $\beta\alpha<1/2$, otherwise $A_\mu^0$ has negative kinetic energy and is ghostlike. From Eq. (\ref{beta-largealpha}) we see that the maximum value of $\beta\alpha$ is $\sim 1/6<1/2$ which occurs for very large values of $|\alpha|$ on the $\beta_+$ branch. Thus we see that the limit is always satisfied and the zero mode has positive energy. 

Notice that in our approach, there are two relations that must be tuned: 1) the bulk and boundary masses must satisfy relation (\ref{tune}) to permit a zero mode solution consistent with boundary conditions, and 2) a precise boundary kinetic term (one on each brane) with a coefficient given by (\ref{beta}), which is related to the boundary mass $\alpha$ must be added to the action.  
Of course, all of these parameters will receive quantum corrections, and therefore these relations are inherently radiatively unstable. This question will not be addressed in this paper and we simply assume that the tuning is enforced in the underlying theory to cancel any quantum effects that leads to deviations in the tunings.

One might hope that there is some symmetry which could enforce the tuning relations. One possibility is supersymmetry. Similar bulk and boundary mass relations for scalar fields are demanded by supersymmetry \cite{gp1}, and it is not unrealistic to think that the same thing could happen for gauge fields. However there are no known examples where boundary kinetic terms are demanded by supersymmetry, so it remains an unanswered question whether the second tuning will be enforced by any symmetry.

\subsubsection{Different boundary kinetic terms}
\begin{figure}
\centerline{\includegraphics[width=.85\textwidth]{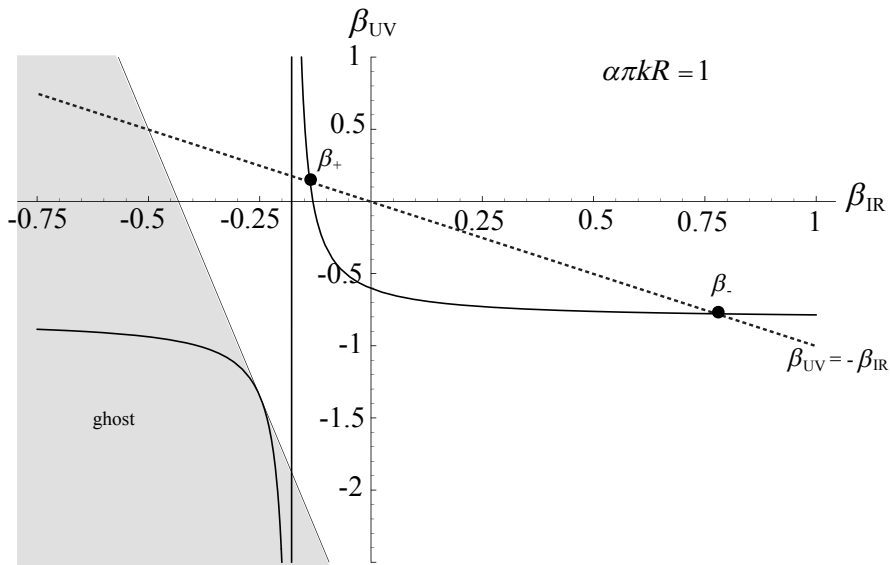}}
\caption{The allowed values of $\beta_{UV}$ and $\beta_{IR}$ for the case $\alpha \pi k R=1$, indicated by the solid line. The shaded region indicates the $\beta$ values which produce a zero mode ghost, and are thus not allowed. The line $\beta_{UV}=-\beta_{IR}$ is also shown (dotted line), and the $\beta_+$ and $\beta_-$ branches are also indicated. }
\label{fig3}
\end{figure}

Let us briefly discuss the possibility of having different kinetic terms on the UV and IR brane, with strengths $\beta_{UV}$ and $\beta_{IR}$. Modifying the action (\ref{a2}) 
\begin{equation}
S=\int d^5x \left[-\frac{1}{4}\left(1+2\frac{\beta_{UV}}{k^2}\delta(y)+2\frac{\beta_{IR}}{k^2}\delta(y-\pi R)\right)F^a_{\mu\nu}F^{\mu\nu a}+\dots\right],
\label{adiff}
\end{equation}
we can follow the procedure outlined in Section 2.2 and compute the zero mode wavefunction and cubic and quartic couplings. Demanding that the couplings are equal gives a constraint on the allowed $\beta$ values.  For a given localization $\alpha$, we can choose, say, $\beta_{IR}$ and then compute $\beta_{UV}$ in terms of $\alpha$ and $\beta_{IR}$. We must also demand that the normalization of the zero mode wavefunction is real;  otherwise the zero mode will be a ghost. This is illustrated in Fig. \ref{fig3} for the case $\alpha \pi k R=1$.
Notice that it is not possible to have two positive kinetic terms; at least one must be negative. This implies that for high energy processes, locally the kinetic term will appear to be negative, and the theory will have a ghost. At large distances, the nonlocal effects of both brane terms combine to form a perfectly sensible effective field theory, so long as the boundary terms are not so strong and negative that they overwhelm the positive bulk contribution. The shaded region indicates where this happens; the negative (and large) brane kinetic terms produce a zero mode ghostlike gauge field. Note also the case $\beta_{UV}=-\beta_{IR}$ studied in the previous section lies in the ghost-free region. 

The structure of this plot is similar for different values of the localization parameter $\alpha$. As we localize the zero mode towards the IR brane (increasing $\alpha$) the asymptote moves towards, but never reaches, the $y$-axis. Instead localizing the massless mode towards the UV brane (decreasing $\alpha$), causes the asymptote to move further away towards large negative values. The qualitative behavior of the allowed values for $\beta_{UV}$ and $\beta_{IR}$ is the same in these cases. In particular, it is never possible to have both brane kinetic terms positive, and there is always a region where the brane kinetic terms produce a zero mode ghost.

\section{The scalar sector}
To localize a massless gauge field in the bulk, we found that it was necessary to add mass terms in the bulk and on the boundary. If we simply add a hard mass term which breaks the 5D gauge symmetry, there is no hope of recovering a 4D gauge symmetry in the effective theory. Thus, it is necessary to generate the mass term through spontaneous symmetry breaking, i.e. a Higgs mechanism. Thus localizing a gauge field in this approach comes at the price of introducing additional scalar fields in the theory. These additional scalars complicate the analysis and obscure the 4D gauge symmetry compared to the case of a massless 5D bulk gauge field, in which the 4D symmetry is realized in a rather simple way. In that case, the scalar mode $A^n_5$ is the Nambu-Goldstone boson which is eaten by the massive vector $A_\mu^n$ \cite{rsch}. With additional scalar fields, the Nambu-Goldstone boson will be a particular linear combination of scalar fields, while another combination will be a physical scalar degree of freedom. Although the analysis is complicated in warped space, we illustrate this feature with a flat space example in Section 3.1. 

More pressing is the possibility that these scalar modes may be massless, strongly coupled, or even ghostlike. We discuss these issues in Section 3.2 using a simple example in warped space with only boundary masses. We find that natural boundary conditions allow a zero mode physical scalar field to exist. To preserve the gauge symmetry a boundary mass term for the vector $A_\mu$ necessarily contains a scalar kinetic term. When the bulk and boundary mass parameters are tuned to cancel the zero mode mass, we might then expect that the zero mode scalar kinetic term is canceled or even becomes negative, corresponding to a strong coupling problem or a ghostlike state, respectively. Although we find no ghost in this example, the physical scalar zero modes are generically strongly coupled. It seems likely that a different choice of boundary conditions (e.g. Dirichlet) will eliminate the scalar zero mode, but the general analysis is then complicated, as we will discuss. In this section our discussion is limited to the Abelian case for simplicity, but of course similar issues will be faced in non-Abelian extensions.

\subsection{Nambu-Goldstone bosons}
Let us revisit the Abelian case, and consider the 5D action for the gauge field $A_M(x,y)$ with bulk mass $m$ and the 5D Nambu-Goldstone boson $\varphi(x,y)$ which appears after spontaneous symmetry breaking:
\begin{equation}
S  = \int d^5x\sqrt{-g}\left[-\frac{1}{4}F_{MN}F^{MN}-\frac{1}{2} m^2\left(A_M- \frac{1}{m} \partial_M \varphi\right)^2 +...\right]~,
\label{a3}
\end{equation}
where we have only included quadratic terms in the Lagrangian involving $A_M$ and $\varphi$. From a 5D point of view, $\varphi$ is the Nambu-Goldstone boson which gives a mass to the vector $A_M$. However when we compactify the extra dimension, the modes $A^n_5$ will play some role in giving the 4D vectors $A^n_\mu$ a mass $m_n$. We might then expect that it is some combination of modes $\varphi^n(x)$  and $A^n_5(x)$ which form the true Nambu-Goldstone boson for $A^n_\mu$, while the orthogonal combination is a physical scalar particle. 

Previously we did not take into account the Nambu-Goldstone boson which couples to $A_\mu$ at the quadratic level in the action. Here we show that it is indeed possible to isolate $A_\mu$ by defining $R_\xi$ gauges. By separating the action (\ref{a3}) into components $\mu, 5$ we obtain
\begin{eqnarray}
S & = & \int d^5x \left[ -\frac{1}{4}F_{\mu\nu}^2-\frac{1}{2}e^{-2 k y}(\partial_5A_\mu)^2-\frac{1}{2}m^2 
e^{-2ky}(A_\mu)^2 \right.\nonumber \\
& &-\frac{1}{2}e^{-2ky}(\partial_\mu A_5)^2 -\frac{1}{2}e^{-4ky}m^2 (A_5)^2\nonumber \\
& & -\frac{1}{2}e^{-2ky}(\partial_\mu \varphi)^2 -\frac{1}{2}e^{-4ky} (\partial_5\varphi)^2\nonumber \\
& & \left.+ e^{-2ky}\partial_\mu A_5 \partial_5 A^\mu +e^{-2ky} m A^\mu \partial_\mu \varphi + e^{-4ky}m A_5 \partial_5 \varphi +...\right]~.
\label{a4}
\end{eqnarray}
We can see that there is nontrivial mixing between $A_\mu, A_5$, and $\varphi$, and it is not clear which combination of fields is the Nambu-Goldstone boson. We can remove the mixing in the $A_\mu$ sector by adding a gauge fixing term
\begin{equation}
{\cal L}_{GF}=-\frac{1}{2\xi}\left[\partial_\mu A^\mu +\xi(\partial_5 e^{-2ky}A_5-e^{-2ky}m\varphi)\right]^2~.
\label{lgf}
\end{equation}
The bulk equation of motion (\ref{eom2}) then follows if we choose the unitary gauge given by $\xi \rightarrow \infty$.
With boundary mass terms there will also be terms involving $\varphi$ in the boundary action that mix with $A_\mu$. This mixing can be removed in a similar manner by gauge fixing, except that in this case there will be terms proportional to delta-functions. Still the vector $A_\mu$ can be isolated and its equation of motion can be analyzed straightforwardly as in Section 2.1. 

Ideally, we would like to decompose the action (\ref{a4}) in order to see how the 5D symmetry manifests itself in the Kaluza-Klein action (i.e. identify the Nambu-Goldstone boson), as is done with massless gauge fields \cite{rsch}. In warped space, however, it is unclear how to decompose the fields since they obey different equations of motion. The analysis can be performed in flat space, and we do find that the 4D Nambu-Goldstone boson is a linear combination of $A^n_5$ and $\varphi^n$.

\subsubsection{Minkowski space analysis}

Consider the flat space limit $k\rightarrow 0$ of the action (\ref{a4}):
\begin{eqnarray}
S & = & \int d^5x \left[ -\frac{1}{4}F_{\mu\nu}^2-\frac{1}{2}(\partial_5A_\mu-\partial_\mu A_5)^2\right. \nonumber \\
& &\left. -\frac{1}{2}m^2\left(A_\mu-\frac{1}{m}\partial_\mu\varphi\right)^2 -\frac{1}{2}m^2 \left(A_5-\frac{1}{m}\partial_5\varphi\right)^2  \right]~.
 \label{flataction}
\end{eqnarray}
The structure of the action gives us a hint as to how to decompose the fields. Consider  the following expansion:
 \begin{eqnarray}
A_\mu(x,y)&=&\sum_{n=0}^\infty A^{n}_\mu(x)f^{n}(y)~, \label{Aex} \\
A_5(x,y)  &=  &\sum_{n=0}^\infty  A^{n}_5(x)\frac{1}{\omega_n}\partial_5 f^{n}(y)~, \label{a5ex}\\
\varphi(x,y) & = & \sum_{n=0}^\infty \varphi^n(x) f^n(y)~, \label{phiex}
\end{eqnarray}
where $(\partial_5^2+\omega_n^2) f^n(y)=0$ and the normalization is determined by 
\begin{equation}
\int_0^{\pi R} dy~f^n(y)f^m(y)=\delta^{nm}.
\end{equation}
The resulting 4D action is diagonal in Kaluza-Klein modes, but there is still mixing between $A^n_\mu, A^n_5$, and $\varphi^n$:
\begin{eqnarray}
S & = & \int d^4x \sum_n \left[ -\frac{1}{4}(F^n_{\mu\nu})^2 -\frac{1}{2}(\omega^2_n+m^2)(A^n_\mu)^2+\omega_nA^n_\mu \partial^\mu A^n_5 +m A^n_\mu \partial^\mu \varphi^n \right.\nonumber \\
& &\left. -\frac{1}{2} (\partial_\mu A^n_5)^2 -\frac{1}{2}m^2 (A^n_5)^2  -\frac{1}{2}(\partial_\mu \varphi^n)^2  -\frac{1}{2}\omega_n^2 (\varphi^{n})^2 +m\omega_n A^n_5 \varphi^n\right]~.
\end{eqnarray}
Examining the action, it is natural to guess that the Nambu-Goldstone boson is proportional to the linear combination 
$\omega_n A^n_5 +m\varphi^n$. However, we must also show that there is a scalar massless eigenstate. The mass mixing matrix
\begin{equation}
 \left( \begin{array}{cc} m^2 & -m \omega_n \\ -m \omega_n &  \omega^2_n \end{array} \right)~,
\end{equation}
has eigenvalues ($0$, $m^2+\omega_n^2$) and the eigenstates can be written as
\begin{eqnarray}
\psi^n=\frac{1}{\sqrt{\omega_n^2+m^2}}(\omega_n A^n_5 + m \varphi^n)~, \nonumber \\
\rho^n=\frac{1}{\sqrt{\omega_n^2+m^2}}(-m A^n_5+\omega_n\varphi^n)~.
\end{eqnarray}
Rewriting the action in terms of scalar mass eigenstates, we find
\begin{eqnarray}
S & = & \int d^4x \sum_n \left[ -\frac{1}{4}(F^n_{\mu\nu})^2 -\frac{1}{2}(\omega^2_n+m^2)(A^n_\mu)^2+ \sqrt{\omega_n^2+m^2} A^n_\mu \partial^\mu \psi^n \right. \nonumber \\
& & \left.-\frac{1}{2} (\partial_\mu \psi^n)^2  -\frac{1}{2}(\partial_\mu \rho^n)^2  -\frac{1}{2}(\omega_n^2+m^2) (\rho^{n})^2 \right], \label{bulk}
\end{eqnarray}
It is clear that $\psi^n$ is the Nambu-Goldstone boson. The coefficient of the mixing term between $A^n_\mu$ and $\psi^n$ is exactly the one needed to make the propagator for $A^n_\mu$ transverse. The unitary gauge is simply $\psi^n=0$. Notice that $\rho^n$ is a physical scalar degree of freedom, as we would expect from counting the degrees of freedom contained in a 5D massive vector field. 
 
If no additional boundary terms are considered, then the low energy theory contains no massless modes. It is possible to obtain massless modes by augmenting the bulk action with boundary potentials which modify the boundary conditions. 

In flat space the Kaluza-Klein decomposition is simple since the eigenfunctions are sinusoidal. In warped space, it is not clear that the eigenfunctions of the vector $A_\mu$ and the physical scalar and Nambu-Goldstone bosons will be related in a simple manner. In the case with no bulk and boundary masses \cite{rsch}, the decomposition proceeds as in (\ref{Aex}) and (\ref{a5ex}) and the 4D action is simple.   

\subsection{Strong coupling and ghosts}
Besides complicating the general analysis, scalar fields can in principle introduce more severe problems into the theory. To add boundary mass terms for the vector, we must also add kinetic terms for the scalar fields on each brane. The tuning condition (\ref{tune}) implies that one of these kinetic terms has the wrong sign, associated with negative energy. At low energies this does not necessarily mean that the scalar has negative kinetic energy, since there are contributions from the bulk and the brane to the kinetic term of the scalar field. One must explicitly examine the low energy theory to see if the particle is a ghost. Indeed, even though negative kinetic energy terms were added for the gauge field in Section 1.2 to maintain 4D gauge invariance, the zero mode did have overall positive energy (\ref{a2}). However
this cancellation mechanism was nonlocal so that at high energies there is a ghost locally which eventually needs to be resolved in the UV completion of the theory.

Another potential problem is that the zero mode scalars might have a small or even vanishing kinetic term due to cancellations from bulk and brane contributions. The kinetic term must then be canonically normalized, leading to strong coupling in any interactions in which the scalar participates. These issues are difficult to examine in the general warped case with bulk and boundary masses because the action (\ref{a4}) does not allow for a straightforward Kaluza-Klein decomposition of the scalar fields. However, there is one warped special case, the ``specular'' solution in which the bulk mass is zero but boundary mass terms localize the gauge field, where we can analyze these scalar issues in detail. We find in this case that although there is no ghostlike particle, there is a strong coupling problem (i.e. the kinetic term vanishes). We give suggestions as to how this problem might be eliminated.

\subsubsection{``Specular'' solution}

A massless gauge field in the bulk of AdS$_5$ is described by the action 
\begin{eqnarray} 
S & = &\int d^5x\sqrt{-g}\left[-\frac{1}{4}F_{MN}F^{MN}\right]~,\nonumber \\
 & = & \int d^5x\left[-\frac{1}{4}F_{\mu\nu}F^{\mu\nu}-\frac{1}{2}e^{-2ky}(\partial_5 A_\mu -\partial_\mu A_5)^2 \right]~.
\end{eqnarray}
Following the standard procedure \cite{rsch}, we expand the 5D fields in the following  way:
\begin{eqnarray}
A_\mu(x,y)&=&\sum_{n=0}^\infty A^{n}_\mu(x)f^{n}(y)~, \\
A_5(x,y)  &=  &\sum_{n=0}^\infty A^{n}_5(x)\frac{1}{m_n}\partial_5 f^{n}(y)~,
\label{exp}
\end{eqnarray}
where the eigenfunctions $f^n(y)$ satisfy the differential equation
\begin{equation}
\partial_5 e^{-2ky}\partial_5 f^n(y) = -m_n^2 f^n(y)~,
\end{equation}
and the orthogonality condition
\begin{equation}
\int_0^{\pi R} dy~f^n(y)f^m(y)=\delta^{nm}~.
\end{equation}

We will focus on the massless mode, $m_0=0$. Normally the solution is taken to be $f^0=constant$,
but since the differential equation is second order, there are in fact two linearly independent solutions. The general solution is given by
\begin{equation}
f^0(y)=C_1+C_2e^{2ky}~.
\end{equation}

We must impose boundary conditions that are consistent with the variational principle. The boundary terms that arise from variation of the action are
\begin{equation}
\delta S =\int d^4x~\delta A^\mu\left[-e^{-2ky}\partial_5 A_\mu +e^{-2ky}\partial_\mu A_5\right]
\bigg\vert^{\pi R}_0~,
\label{b1}
\end{equation}
which forces us to impose Dirichlet or Neumann boundary conditions on $f^0$. These choices force $C_2=0$, and the exponential solution is removed. However, it is possible to keep the exponential solution if we add the following boundary mass term to the action:
\begin{equation}
S_{bdy}=-\int d^5x \sqrt{-g}\left[ 2k \left( A_\mu -\frac{1}{2k}\partial_\mu A_5\right)^2(\delta(y)-\delta(y-\pi R))\right]~.
\label{bmass}
\end{equation}
We will see that the boundary action is invariant under a 4D gauge symmetry. Varying the boundary action yields 
\begin{eqnarray}
\delta S_{bdy} & =& \int d^4x\left[ \delta A^\mu\left( e^{-2ky}2k A_\mu -e^{-2ky}\partial_\mu A_5\right)\right.\nonumber \\
&&\left. + ~\delta A_5 \left(-\frac{1}{2k}e^{-2ky}\Box A_5 + e^{-2ky}\partial \cdot A\right)\right]
\bigg\vert^{\pi R}_0~.
\label{b2}
\end{eqnarray}
Combining the terms in (\ref{b1}) and (\ref{b2}) and using the bulk equation of motion for $A_5$:
\begin{equation}
\Box A_5-\partial_5 \partial\cdot A=0~,
\end{equation}
we find the natural boundary condition for $f^n(y)$ to be
\begin{equation}
(\partial_5-2k)f^n\bigg\vert_{0, \pi R} = 0~.
\label{bc3}
\end{equation}
This corresponds to $\alpha=2$ in (\ref{bc1}), which is consistent with (\ref{tune}). Notice that there are other possibilities for boundary conditions consistent with the variational principle. For instance we can choose Dirichlet conditions for $A_5$. In this case the analysis is not as straightforward and the 4D gauge invariance is obscured, but there will be no $A_5$ zero mode. We will revisit this discussion shortly when we discuss strong coupling.

For the zero mode, the exponential solution satisfies the boundary condition (\ref{bc3}) and the constant $C_1=0$. The zero mode solution is given by 
\begin{equation}
f^0(y)=\sqrt{\frac{4k}{e^{4\pi k R} -1}}e^{2ky}~.
\end{equation}

Note that the zero mode field $A^0_5(x)$ is not eliminated in this case, but has the same bulk profile as $A^0_\mu$ in the fundamental domain $y\in(0,\pi R)$. This is consistent because the wavefunction for $A^0_5$ is $\sim {\rm sgn}(y)e^{2ky}$ and is therefore odd with respect to orbifold symmetry.  

As mentioned above, the boundary action (\ref{bmass}) respects a 4D gauge invariance:
\begin{eqnarray}
\delta A_\mu & = & \partial_\mu \lambda~, \\
\delta A_5 &=&2k \lambda~,
\end{eqnarray}
where, noting (\ref{bc3}), the second relation is equivalent to $\delta A_5=\partial_5\lambda$ on the boundary.

It is instructive to examine how the {\it Abelian} gauge symmetry, which involves no Nambu-Goldstone fields, arises for the zero mode. To this end, we will detune the condition (\ref{bc3}) to be 
\begin{equation}
(\partial_5 -(2- \epsilon) k)f^n\bigg\vert_{0,\pi R}=0~,
\label{bc4}
\end{equation}
where $\epsilon$ is now an arbitrary constant. The lowest lying mode becomes massive. The eigenfunctions are given by 
\begin{equation}
f^n(y)=N_n e^{ky}\left[J_1\left(\frac{m_n}{k e^{-ky}}\right)+ \kappa(m_n) Y_1\left( \frac{m_n}{k e^{-ky}}\right)\right]~,
\end{equation}
where $N_n$ and $\kappa$ are constants. The coefficient $\kappa$ is found by applying the boundary conditions:
\begin{equation}
\kappa(m_n)=-\frac{\epsilon J_1\left(\frac{m_n}{k}\right)-\frac{m_n}{k}J_2\left(\frac{m_n}{k}\right)}{\epsilon Y_1\left(\frac{m_n}{k}\right)-\frac{m_n}{k}Y_2\left(\frac{m_n}{k}\right)}~,
\end{equation}
and the masses are found by equating $\kappa(m_n)=\kappa(m_n e^{\pi kR})$.

We will now show that there is a smooth limit $m_0\rightarrow0$ starting from the detuned case. Performing the Kaluza-Klein reduction, the action becomes
\begin{eqnarray} 
S& =& \int d^4x\sum_n\left[-\frac{1}{4}(F^n_{\mu\nu})^2-\frac{1}{2}m_n^2\left(A^n_\mu -\frac{1}{m_n}\partial_\mu A^n_5\right)^2 \right]~, \nonumber \\
&=& \int d^4x\sum_n\left[-\frac{1}{4}(F^n_{\mu\nu})^2-\frac{1}{2}(\partial_\mu A^n_5)^2-\frac{1}{2}m_n^2(A^n_\mu)^2+m_n A^n_\mu \partial_\mu A^n_5 \right]~.
\label{a8}
\end{eqnarray}
It is clear from the action that $A^n_5$ is the Nambu-Goldstone boson, which provides a correct contribution to the vacuum polarization amplitude of $A^n_\mu$. Expanding the gauge parameter $\lambda(x,y)=\sum_n\lambda^n(x)f^n(y)$, we see that the 4D gauge invariance is given by
\begin{eqnarray}
A^n_\mu &\rightarrow& A^n_\mu +\partial_\mu \lambda^n~,\\
A^n_5 &\rightarrow& A^n_5+m_n\lambda^n~.
\label{gs}
\end{eqnarray}
In the unitary gauge, $A_5=0$, the Nambu-Goldstone $A^n_5$ is eaten by $A^n_\mu$. 

Now if we take the limit $\epsilon \rightarrow 0$, the zero mode becomes massless $m_0\rightarrow 0$. The zero mode action reduces to 
\begin{equation}
\int d^4x \left[-\frac{1}{4}(F^0_{\mu\nu})^2-\frac{1}{2}(\partial_\mu A^0_5)^2\right]~,
\end{equation}
and there is an Abelian gauge invariance, as is evident from (\ref{gs}) by taking $m_0=0$. It is also clear from (\ref{gs}) that when $m_n\rightarrow 0$ that we do not have freedom to set the zero mode $A_5^0=0$. 

\subsubsection{Strong coupling}
The expansion (\ref{exp}) indicates that in the limit $m_0\rightarrow0$, the wavefunction for $A^0_5$ becomes infinite. We might argue that the expansion is not valid in this limit and instead use the following expansion:
\begin{equation}
A_5(x,y)  =  \sum_n   A^{n}_5(x)\partial_5 f^{n}(y)~.
\label{exp2}
\end{equation}
In this case we would find that although the wavefunction is finite, the kinetic term for $A^0_5$ vanishes. Actually, this implies that as we tune the boundary mass parameter $\epsilon \rightarrow 0$ in  (\ref{bc4}) to get a massless mode, we should properly canonically normalize the kinetic term. The smallness of the kinetic term then reappears as the large wavefunction in (\ref{exp}).
 
The large wavefunction in (\ref{exp}) means that any interactions will become strongly coupled, and therefore the theory is not well-defined if such interactions exist. Indeed, any bulk interactions with 
$A_M$ would have this problem because of 5D gauge invariance. The problem is even more severe because the mode is massless, meaning that the strong coupling problem is present in the low energy theory. However there are no ghosts in the theory at low energies, as can be seen clearly by taking the limit $m_0\rightarrow0$ in the action (\ref{a8}). The kinetic term for $A_5$ is always properly normalized with the correct sign corresponding to positive energy. 

How can we avoid this strong coupling problem? The most obvious possibility would be to impose a different boundary condition for $A_5$ consistent with the variational principle. A Dirichlet condition, 
\begin{equation}
A_5\big\vert_{0,\pi R}=0~,
\end{equation}
is our other option consistent with the variational principle, as can be seen from examining (\ref{b2}).
This implies $\delta A_5=0$ on the boundary and will make the second term in (\ref{b2}) vanish. More importantly, this boundary condition will kill the zero mode $A_5^0$. At the massless level, we would have an ordinary gauge theory with only the zero mode $A_\mu^0$. However, for the massive modes, the action would no longer be diagonal with respect to Kaluza-Klein level $n$ because the wavefunctions associated with $A_5^n(x)$ will be altered. In particular the zero mode would mix with these massive modes quadratically and the 4D gauge symmetry would be obscured at high energies. In principle there is no problem with this philosophy, and in fact this is akin to what happens in non-Abelian theories when the zero mode interacts with the Kaluza-Klein modes via cubic and quartic terms, and the gauge invariance is realized in a non-trivial way (each Kaluza-Klein mode transforms into other Kaluza-Klein modes). 

Another way the strong coupling problem might be addressed is via quantum corrections. In the $m_n\rightarrow 0$ limit, the mass of the $A_5^0$ is not protected by a gauge symmetry in the 4D theory (at least in the Abelian case), and thus we would expect the mass to receive large corrections, quadratically dependent on the UV cutoff of the theory. In this way $A_5^0$ could decouple at low energies, relegating the problem to the UV. On the other hand, the masslessness of $A^0_\mu$ is protected by 4D gauge symmetry. 
 
\section{Other approaches to gauge field localization}

We have seen that bulk and boundary masses alter the equations of motion and boundary conditions to permit a localized zero mode non-Abelian gauge field. However, this is not the only way to achieve localization. Another possibility is to couple the gauge field to a dilaton with a nontrivial background solution, as first illustrated in \cite{dilaton1}. Interestingly, it has been suggested that this approach is related to localization via bulk and boundary mass terms, essentially through a field redefinition \cite{dilaton3}. This connection has also been alluded to via deconstruction, in which the wavefunction of the zero mode results from a position dependent gauge coupling varying from site to site \cite{decon}. In this section, we wish to comment on this relation and point out the similarities and differences in each approach. In the end, these other approaches to gauge field localization may provide simpler models, in particular with regard to the scalar sector, which we have seen in our approach is quite difficult to analyze in a satisfactory way.

\subsection{Localizing the gauge field with a dilaton}
Let us begin by reviewing the basic idea considered in \cite{dilaton1}. The theory consists of 5D gravity coupled to a  scalar field $\phi$. The scalar $\phi$ acquires a bounce-like configuration, which can be idealized to a ``brane'' by taking a particular limit in the scalar potential. RS2-like solutions appear for the metric in this brane limit. 

If we introduce another scalar, the dilaton $\pi$, it also acquires a nontrivial background, which in the brane limit turns out to be linear in $y$, $\pi(y)=b k y$, where $b$ is a dimensionless constant. In fact if we simply begin with a brane, that is, without reference to the field $\phi$, we still find a linear dilaton solution \cite{dilaton1}. In these cases, however, the dilaton $\pi$ back-reacts on the geometry, which is modified from pure AdS$_5$. In our discussion below, we will assume the geometry is pure AdS$_5$ to compare with our model, even though we know of no 5D model with a linear dilaton solution and pure AdS$_5$ geometry.   
     
Following \cite{dilaton1}, we can easily see how the gauge field is localized. Consider the following action of a massless gauge field coupled to a dilaton $\pi$, expanded about its background configuration:
\begin{eqnarray}
S &=& \int d^5 x \sqrt{-g}  \left[-\frac{1}{4}e^{-2\pi}F_{MN}F^{MN}~\right]~,\nonumber \\
&=& \int d^5x~ e^{-2 b k y}\left[-\frac{1}{4}F_{\mu\nu}F^{\mu\nu}-\frac{1}{2}e^{-2 k y} (\partial_5 A_\mu-\partial_\mu A_5)^2 \right]~.
\label{a5}
\end{eqnarray}
Performing a Kaluza-Klein decomposition as in (\ref{kk}), the solution for the massless mode is a constant, $f^0(y)=N$. The zero mode action can then be written as 
\begin{equation}
N^2\int_0^{\infty} dy~ e^{-2 b k y} \int d^4x\left[ -\frac{1}{4}F^0_{\mu\nu}F^{\mu\nu 0} ~\right]~.
\end{equation}
The profile of the gauge boson with respect to a flat metric is then seen to be $\tilde{f}^0(y) \sim e^{-b k y}$. The coupling to the dilaton simply becomes the gauge boson wavefunction when the dilaton acquires a vacuum expectation value (VEV).

This model can be related to localization with bulk and boundary masses by performing a field redefinition, $A_M\rightarrow e^{bky}A_M$. The action (\ref{a5}) transforms to 
\begin{equation}
S = \int d^5x~ \left[-\frac{1}{4}F_{\mu\nu}F^{\mu\nu}-\frac{1}{2}e^{-2 k y} (\partial_5 A_\mu +bk A_\mu-\partial_\mu A_5)^2 \right]~.
\label{a6}
\end{equation}
It appears that we have broken the 5D gauge symmetry by performing this $y$-dependent redefinition, but of course the symmetry is manifested in a different way. Gauge transformations on the vector $A_\mu$ are the same while transformations on $A_5$ become rather unusual. Under field redefinition, the gauge parameter transforms to $\lambda \rightarrow e^{bky}\lambda$, and we find
\begin{eqnarray}
A_\mu & \rightarrow &A_\mu+\partial_\mu \lambda~, \\
A_5 & \rightarrow & A_5+\partial_5 \lambda +bk\lambda~.
\label{gauge}
\end{eqnarray}

We can still choose the gauge $A_5=0$\footnote{Indeed, we can choose this gauge prior to redefining the field in (\ref{a5}).}, in which case the action (\ref{a6}) can be rewritten as
\begin{eqnarray}
S &= &\int d^5x~ \left[-\frac{1}{4}F_{\mu\nu}F^{\mu\nu}-\frac{1}{2}e^{-2 k y} (\partial_5 A_\mu)^2 -\frac{1}{2}(b^2+2b) k^2e^{-2 k y}(A_\mu)^2\right]\nonumber \\
& & +\int d^4x~\frac{1}{2}b k\, e^{-2 k y} (A_\mu)^2 \bigg\vert_0~.
\label{a7}
\end{eqnarray}
We see that the field redefinition has generated bulk and boundary mass terms. The relation of the coefficients of these mass terms is precisely the one necessary to force a localized massless mode, which can easily be seen by identifying $b=-\alpha$ and $a=b^2+2b$, in which case the tuning relation (\ref{tune}) holds. Again the wavefunction is given by $\tilde{f}^0(y)\sim e^{-b k y}$. Notice that in this model \cite{dilaton1}, there is only one brane, so that the wavefunction is normalizable only if $b$ is positive. 

Why can't we simply start from the action (\ref{a6}), and demand the strange gauge symmetry (\ref{gauge}) without ever referring to the dilaton $\pi$? This would allow us to have a theory with a localized gauge field and {\it no} extra scalar fields. We can do this, but there are a couple of objections to this approach. The action ({\ref{a6}}) explicitly violates 5D Lorentz invariance in the bulk. Of course the dilatonic action (\ref{a5}) appears to violate the symmetry as well, but this is only because the VEV of the dilaton spontaneously breaks Lorentz invariance, and the symmetry is therefore realized non-linearly through fluctuations about the background. If we don't have this extra scalar, then the theory breaks 5D Lorentz invariance. On the other hand, 4D Lorentz invariance is a symmetry of the action (\ref{a6}), which is the only symmetry we need for phenomenological purposes. Another objection, of course, would be that the unorthodox gauge symmetry in (\ref{gauge}) would seem a bit unnatural if we didn't know about its origin from the dilaton coupling. It is a local symmetry nonetheless, indicating a redundancy in the description of the physical degrees of freedom in the system.

\subsection{Localizing the gauge field via deconstruction}
The pure action (\ref{a6}) without a dilaton finds a stronger motivation through deconstruction as shown in \cite{decon}. In this section we will review this approach and mention some of the advantages of the deconstruction viewpoint. Consider $N$ sites each with an SU$(n)_i$ gauge theory. The sites are connected by bifundamental scalar fields $Q_i$. The 4D Lagrangian is then
\begin{equation}
{\cal L}=-\sum_{n=1}^{N-1}(D_\mu Q_i)^\dag (D^\mu Q_i)~,
\end{equation}
where the covariant derivative is defined as $D_\mu Q_i=\partial_\mu Q_i-ig_iA^i_\mu Q_i+ i g_{i+1}Q_iA_\mu^{i+1}$. The gauge symmetry is broken to a diagonal SU$(n)_{diag}$ by assuming that the scalar fields acquire VEVs proportional to the unit matrix, $\langle Q_i \rangle = v_i/\sqrt{2}$. The mass Lagrangian is then
\begin{equation}
{\cal L}_m=-\frac{1}{2}\sum_{i=1}^{N-1}|v_i(g_{i+1}A_\mu^{i+1}-g_i A_\mu^i)|^2~.
\end{equation}
To reproduce the familiar case of a massless gauge field in AdS$_5$ in the continuum limit, the prescription is to take universal gauge couplings $g_i=g$ and $y$-dependent VEVs $v_i=v e^{-ki/(gv)}$, where the site ``coordinate'' is $y=i/(gv)$. This gives a flat zero mode gauge field. However, if we make different choices for the couplings and VEVs, in particular assuming nonuniversal gauge couplings of the form $g_i=g e^{bki/(gv)}$ and VEVs of the form $v_i=v e^{-(1+b)ki/(gv)}$ the continuum limit mass Lagrangian becomes
 \begin{equation}
{\cal L}_m=\int_0^{\pi R} dy\left[-\frac{1}{2} e^{-(2+b)ky}\partial_5\left(e^{-2 b k y}A_\mu(x,y)\right)^2\right]~.
\end{equation}
This action can be rewritten identically to (\ref{a7}), containing the precise bulk and boundary mass terms which will localize a massless mode. From the deconstructed point of view, this makes complete sense because the SU$(n)_{diag}$ gauge symmetry is still intact, implying the existence of a massless eigenstate which is a linear combination of the $A^i_\mu$. What is appealing about the deconstruction approach to localizing the gauge field is that there is no reference to a dilaton. The fluctuations about the vacuum $v_i$ are related in the continuum  to $A_5$, and no other scalar field is present in the theory. Indeed, after taking into account these fluctuations, the full action is given by (\ref{a6}). The zero mode localization is simply a consequence of different gauge theories $A_\mu^i$ having different couplings $g_i$, which is certainly allowed. The absence of the dilaton is much like the absence of the radion in generic deconstructed theories compared to their full five dimensional counterparts. 

\subsection{Comparison}
The approaches to localizing gauge fields described in the previous section seem very similar to our approach. The localized wavefunction can be ascribed to a dilaton coupling in a nontrivial background, or to a position dependent gauge coupling. Upon a field redefinition, bulk and boundary masses appear with precisely the right coefficients necessary to localize a zero mode, identical to (\ref{mass}). These observations were first made in \cite{dilaton3}.

Although the technical details of each approach are similar, the philosophy behind each is somewhat different. We are beginning with a pure gauge theory in a slice of AdS$_5$, and attempting to generate the necessary mass terms through the spontaneous breaking of the symmetry. The 5D bulk Lorentz symmetry is never broken. In the other approaches, the localized wavefunction is generated through an auxiliary mechanism, either the dilaton profile or a position dependent coupling, and is crucially dependent on breaking 5D Lorentz invariance in the bulk. Moreover, it remains to be shown whether there exists a solution to Einstein's equations that produces a linear dilaton background with pure 
AdS$_5$ geometry.

The number of additional physical scalar particles in each theory is different as well. In our approach, the number of scalar fields depends on the representation of the symmetry group in which scalar fields reside, and this number could be quite large. The analysis of these scalars is also complicated since there is mixing at the quadratic level in the action, and can in principle lead to strong coupling problems in the theory, as elaborated on in Section 3. On the other hand, in order to localize the field with a dilaton, only one scalar is needed per gauge field $A_M$. Fluctuations about the dilaton background in (\ref{a5}) need to be studied, but the analysis should be simpler because these perturbations will not mix with the gauge field at the quadratic level, but only appear as interactions. On the other hand, the deconstructed approach is even simpler because there are no additional scalar fields, although the continuum theory explicitly breaks 5D Lorentz symmetry. 
 
At the non-Abelian level, these approaches really diverge. We found it necessary to add brane-localized kinetic terms to restore the equality of the zero mode gauge couplings. The other approaches accomplish this by having position dependent couplings. To see this, consider the non-Abelian generalization of (\ref{a5}). To compare with our approach, we redefine the field $A^a_M\rightarrow e^{bky}A^a_M$ to generate bulk and boundary mass terms. Upon this redefinition, the cubic and quartic interactions become 
\begin{equation}
S_{int}=- \int d^5x \left[g_5 e^{bky} f^{abc}\partial_\mu A^a_\nu A^{\mu b}A^{\nu c}
+\frac{g_5^2}{4}e^{2bky}f^{abc}f^{ade}A^b_\mu A^c_\nu A^{\mu d}A^{\nu e}\right]~.
\label{int2}
\end{equation}
We can view the 5D gauge couplings as position dependent: $\tilde{g}_5=g_5 e^{bky}$. Inserting the zero mode wavefunction $f^0(y)=N e^{-bky}$ and computing the overlap integrals, we find that the 4D cubic and quartic couplings are equal, and thus the low energy theory is gauge invariant. In our approach, the 5D gauge couplings are constant in the bulk. 

The phenomenology of each localization mechanism would also be different. In our approach, brane kinetic terms must be included, which can drastically alter the mass spectrum, wavefunctions, and couplings to matter, as illustrated by a number of studies \cite{blkt1, blkt2}. Brane kinetic terms are not necessary in the other approaches, but of course may be included for study. Finally, the scalar content of each theory is much different, which would affect the phenomenology.

\section{Localizing $A_5$}

Although we have focused on the gauge field $A_\mu$ in this paper, many models utilize the $A_5$ field in one way or another. One well known example is the Hosotani mechanism \cite{hosotani}, or radiative symmetry breaking, where quantum corrections induce a mass for the zero-mode $A^0_5$, which is then identified as the Higgs boson. A recent twist on this idea, motivated by holography, are composite Higgs models \cite{hpgb,mchm}, where the dual of the $A_5$ is a composite pseudo-Nambu-Goldstone boson of a strongly coupled theory. The ability to localize the $A_5$ field in these contexts could have interesting effects on the phenomenology of these models. 
 
With a massless 5D gauge boson in the bulk, the profile of $A^0_5$ is fixed by gauge invariance, and in fact is localized near the IR brane. As is the case with the vector $A_\mu$, it is possible to localize the scalar $A_5$ zero mode at different points in the bulk. Using bulk and boundary masses generated by spontaneous symmetry breaking to localize this field goes against the philosophy of the gauge-Higgs models where $A_5$ is to be used as an {\it alternative} to spontaneous symmetry breaking. Hence, we will show how the $A_5$ can be localized either through a dilaton coupling, or through deconstruction, similar to the case for $A_\mu$ in Section 4. We reiterate that at this point it is unknown whether a 5D system with a linear dilaton solution in a pure AdS$_5$ background exists. 

Starting from the action (\ref{a5}), we add the following gauge fixing term to diagonalize the action:
\begin{equation}
{\cal L}_{GF}=-\frac{1}{2\xi}\left[e^{-bky}\partial_\mu A^\mu +\xi e^{bky}(\partial_5 e^{-2(b+1)ky}A_5)\right]^2~.
\label{lgf2}
\end{equation}
The $A_5$ action becomes
\begin{equation}
S(A_5)=\int d^5 x \left[-\frac{1}{2}e^{-2(b+1)ky}(\partial_\mu A_5)^2-\frac{\xi}2e^{2b k y}\left(\partial_5 e^{-2 (b+1)k y }A_5\right)^2\right]~.
\label{a9}
\end{equation}
Decomposing the field, $A_5(x,y)=\sum_n A^n_5(x) g^n(y)$, we find that the Kaluza-Klein modes have a $\xi$-dependent mass, indicating that they are not physical particles. However for the zero mode this is not the case. The eigenfunction $g^0(y)$ obeys the equation of motion 
\begin{equation}
\partial_5 e^{2 b k y}\partial_5 e^{-2(b+1)ky} g^0(y) = 0~,
\end{equation}
and with the boundary condition $(\partial_5-2(b+1)k)g^0(y) |_{0,\pi R}=0$ has solution
\begin{equation}
g^0(y)= N e^{2(b+1)ky}~.
\end{equation}
Inserting the solution into the action (\ref{a9}), we have
\begin{equation}
S(A_5)=\int dy ~e^{2(b+1) ky}\int d^4x\left[-\frac{1}{2}( \partial_\mu A_5^0)^2~\right]~,
\end{equation}
indicating that with respect to a flat metric, the wavefunction is $\tilde{g}^0(y)\sim e^{(1+b)ky}$.
The physical zero mode is localized on the UV brane when $b<-1$ while for $b>-1$ it is localized on the IR brane. Note that this agrees with the zero mode localization of a pure bulk scalar with bulk mass $b^2-4$ and boundary mass $b+2$~\cite{gp1}.

\section{Discussion and conclusion}
Zero mode gauge fields in a compact extra dimension can be localized through the addition of bulk and brane mass terms. The inclusion of these terms modifies the bulk equation of motion and the boundary conditions such that a zero mode with an exponential wavefunction exists in the theory. In the non-Abelian case, however, the cubic and quartic couplings of this zero mode are not equal, meaning that the gauge symmetry of the 4D theory is lost. We have shown how to recover the 4D gauge symmetry 
by adding gauge interactions on the boundaries. In this way the sum of the bulk plus boundary contributions to the interactions of the zero mode restores the equality of the gauge couplings, and then the 4D gauge invariance appears in its usual form. We can add these brane interactions in a gauge invariant way via brane-localized kinetic terms. One caveat in this approach is that there are two relations in this theory that must be tuned, one between the bulk and boundary masses to produce a localized massless mode wavefunction, and another between kinetic terms and the boundary mass parameter.  Furthermore even though the zero mode had overall positive kinetic energy, via a nonlocal cancellation mechanism, negative kinetic energy terms were required on at least one brane. These terms would need to be resolved in any sensible UV completion of the model.

There are also caveats related to the additional scalar fields that must be present in the theory.
To add mass terms in the bulk and on the boundary in a gauge invariant way, the terms must be generated by breaking the 5D gauge symmetry, which necessitates introducing additional scalar fields in the theory. In the warped case, a complete analysis of these extra scalars is limited because of the nontrivial mixing. We have instead presented this analysis in flat space, where the eigenfunctions of the different fields are similar. If one could extend this analysis to the warped case, many of the scalar issues, such as strong coupling, could be addressed more generally. We have shown, however, that in the warped case with only boundary masses, a strong coupling problem may exist in the scalar sector. This problem may be resolved by a different choice of boundary conditions or by radiative corrections to the scalar zero mode mass.  

Even if the problems of the scalar sector cannot be addressed without a more fundamental UV description, the Yang-Mills localization will still be preserved at low energies. The phenomenology of localized gauge fields in the Randall-Sundrum model would therefore be interesting to explore. If we consider the Standard Model in the bulk, then each gauge boson can in principle be localized at different points in the extra dimension. Gauge fields in the bulk were originally considered 
in \cite{gaugeb1,gaugeb2}, and it was found that the mass of the first Kaluza-Klein excitation must be pushed to a rather high scale to suppress four-fermion operators. However, localizing gauge fields can help lower the Kaluza-Klein mass scale, as illustrated in the Abelian case \cite{u1}. It is conceivable that localizing gauge fields could have other phenomenological benefits. Also, the phenomenology would be different depending on the localization mechanism, either bulk and boundary masses, dilaton coupling, or deconstruction.  

Perhaps the most interesting aspect of localized gauge fields is the holographic interpretation \cite{u1}. 
If the zero mode is localized on the UV brane, we find that the massless particle in the dual theory is primarily an elementary field external to the CFT. On the other hand, a zero mode localized on the IR brane corresponds in the dual theory to a composite massless gauge field built out of CFT states. For example, if we consider the Standard Model gauge fields propagating in the bulk with zero modes localized towards the IR brane, then in the dual theory the forces that we experience - electromagnetic, weak, and strong - are really emergent phenomena, a result of strongly coupled (unknown) CFT dynamics at low energy. This is similar to the emergent gravity scenario considered in Ref. \cite{gpp}, and adding boundary kinetic terms for gravity as done for non-Abelian gauge fields in this paper, suggests a similar nonlinear extension for gravity.

Finally, we discussed the localization of the $A_5$ field, the scalar component of the 5D gauge field. We have shown that it is possible to localize this field anywhere in the bulk with a dilaton coupling or via deconstruction. This opens up new phenomenological possibilities in models where $A_5$ is the pseudo-Nambu-Goldstone boson responsible for electroweak symmetry breaking. In particular, motivated by the dual interpretation, the dual Higgs field can become even more composite as the zero mode is localized on the IR brane.

\section*{Acknowledgements}
We thank Emilian Dudas and Erich Poppitz for helpful discussions. T.G. acknowledges the CERN Theory Division and the Galileo Galilei Institute for Theoretical Physics in Florence for their warm hospitality and the INFN for partial support where portions of this work were
done. This work was supported in part by a Department of Energy grant DE-FG02-94ER40823 at the University of Minnesota, and an award from Research Corporation.

\end{document}